\documentclass[aps,prl, footinbib,twocolumn,showpacs,amsmath,amssymb,superscriptaddress]{revtex4-2}
\usepackage{graphicx,amsmath}
\usepackage{epstopdf}
\usepackage{amssymb}
\usepackage{grffile}
\usepackage[usenames]{color}
\usepackage{indentfirst}
\usepackage{float}
\usepackage{color}
\usepackage{mathrsfs}
\usepackage{dcolumn}
\usepackage{bm}
\usepackage[colorlinks=true,bookmarks=false,citecolor=blue,linkcolor=red,urlcolor=blue]{hyperref}

\begin{document}

\title{Meissner-Like  Currents of Photons in Anomalous Superradiant Phases}
\author{Linjun Li}
\thanks{These two authors contributed equally}
\affiliation{Department of Physics, Chongqing Key Laboratory for strongly coupled Physics, Chongqing University, Chongqing 401330, China}
\author{Pengfei Huang}
\thanks{These two authors contributed equally}
\affiliation{Department of Physics, Chongqing Key Laboratory for strongly coupled Physics, Chongqing University, Chongqing 401330, China}
\author{Zi-Xiang Hu}
\email{zxhu@cqu.edu.cn}
\affiliation{Department of Physics, Chongqing Key Laboratory for strongly coupled Physics, Chongqing University, Chongqing 401330, China}
\author{Yu-Yu Zhang}
\email{yuyuzh@cqu.edu.cn}
\affiliation{Department of Physics, Chongqing Key Laboratory for strongly coupled Physics, Chongqing University, Chongqing 401330, China}

\begin{abstract}
We present Meissner-like photon currents in a quantum Rabi zigzag chain under staggered synthetic magnetic fields. The ground state of the Meissner superradiant phase hosts persistent chiral edge currents in a sequence of cancellation of antiparallel vortex pairs, akin to surface currents of the Meissner effect in superconductors. The Meissner phase displays distinct vortex structures and anomalous scaling exponents, arising from geometric frustration effects. Modifying the staggered flux triggers transitions to even- or odd-vortex superradiant phases, where the chiral edge currents flow exclusively in even or odd cavities with localized vortices, respectively. Enhanced interspecies interactions induce the vanishing of currents in a ferromagnetic superradiant phase. Our results enable observation of stabilized photon vortices and edge currents with analogy to quantum Hall-like robustness in light-matter coupling systems. 
\end{abstract}

\date{\today }
\maketitle

\textit{Introduction --} 
Remarkable achievements have been made in simulating quantum many-body phenomena via synthetic gauge fields in ultracold atoms and bosonic gases~\cite{RevModPhys.83.1523,RevModPhys.80.885,science2015,PhysRevResearch.1.033102}, akin to charged particles in magnetic fields. By exploiting light-matter interactions, 
optical platforms such as the cavity and circuit QED facilitate realizing  complex many-body interactions with tunability~\cite{science361,nature472,science333,PhysRevLett.130.173601}, leading to exotic quantum phases of matter~\cite{PhysRevLett24xiang,nphys462,PhysRevLett.124.073602,PhysRevLett.132.193602,PhysRevLett.132.053601,NewJPhys,SciChinaPhys}. The superradiant phase transition~\cite{PhysRev.93.99,PhysRevLett.92.073602,PhysRevA78}, a well-known phenomenon in strongly coupled light-matter interactions, has been realized in Bose-Einstein condensates~\cite{nature2010} and Fermi gas~\cite{science2021,PhysRevLett.112.143004}. The quantum Rabi model, as a fundamental light-atom coupling, is recognized to exhibit the superradiant transition~\cite{Ashhab2013,hwang2015,liu2017,chen2020,chen2021,NCcai2021}. In an artificial magnetic field, novel quantum phase transitions are observed, such as chiral superradiant phases in a quantum Rabi triangle~\cite{PhysRevLett.127.063602},  chiral magnetic phases in a quantum Rabi ring~\cite{PhysRevLett.129.183602,PhysRevA.108.043705}, a frustrated superradiant phase~\cite{PhysRevResearch.5.L042016,PhysRevA.110.013713,PhysRevLett.128.163601}, and fractional quantum Hall physics in the Jaynes-Cummnings Hubbard lattice~\cite{PhysRevLett.108.223602,PhysRevA.93.023828}. These advancements demonstrate significant progress in simulating fascinating physics through artificial magnetic fields.

The Meissner effect is an intriguing phenomenon characterizing a superconductor exposed to a magnetic field, where circulating surface currents produce a counteracting field to cancel the applied ﬁeld~\cite{PhysRev1957}. In low-dimensional quantum systems, it has been a long challenge to explore analogous concepts and study the interplay of the magnetic field and many-body interactions. Recently, the distinctive quantum behavior known as chiral Meissner currents, observed in multiple species of neutral particles, has been simulated within two-dimensional bosonic ladders \cite{NaturePhysics588,PhysRevLett111,PhysRevB64} and two-leg ladders \cite{PhysRevLett.132.130601} subjected to an artificial magnetic field. The Meissner currents arise due to the interspecies coherence of interacting bosons. The analogy to the Meissner effect is characterized by chiral edge currents~\cite{NaturePhysics588,PhysRevLett111}, where the intraspecies currents cancel in the bulk but persist at the boundaries, resulting in parallel edge currents along the edges. The chiral edge currents were also realized in zigzag optical ladders \cite{PhysRevLett.124.140401,PhysRevLett1112, PhysRevLett.122.023601,PhysRevA.94.063632}  subjected to synthetic magnetic fields, akin to topological quantum Hall insulators~\cite{PhysRevA.89.023619}.  However,  the technical difficulty in engineering many-body systems and magnetic fields of neutral particles has constrained studies for understanding of analogue ideas of the Meissner effect due to intractability. 

\begin{figure}[tbp]
\includegraphics[scale=0.16]{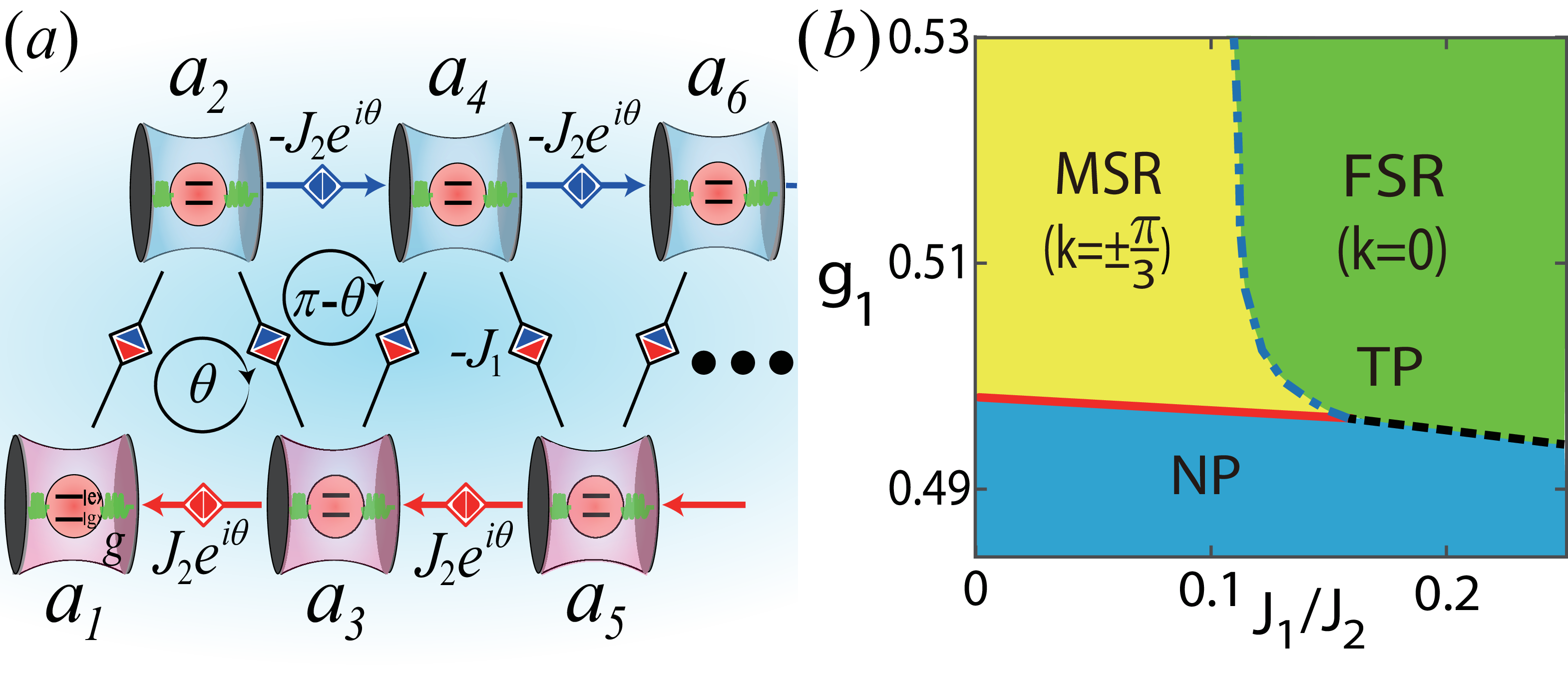}
\caption{(a) Constructing a quantum Rabi zigzag chain exposed to a staggered flux $\theta$ by rearranging a $1$D representation. (b) Phase diagram in ($J_1/J_2$, $g_1$) for $N=6$ cavities with $\theta=\pi/2$. Critical boundary $g_{1c}(k=\pm\pi/3)$ (red solid) and $g_{1c}(k=0)$ (black dash) in Eq. (\ref{gc}) mark the NP-MSR and NP-FSR transitions of second order. First-order line (blue dash dot) converges with the second-order line at a triple point (TP). In all calculations, we set $\omega=1$ as the unit for frequency, and $\Delta/\omega= 50, J_2/\omega= 0.05$.}
\label{model}
\end{figure}

 Considering highly tunable light-matter interactions, we investigate the Meissner-like effect in a specific quantum Rabi chain. Tuning the staggered flux drives transitions to distinct vortex superradiant phases. A novel Meissner superradiant phase exhibits species-specific chiral edge currents and globally canceling antiparallel vortex pairs, analogous to the behavior in the superconductor. Additionally, geometric frustrations give rise to distinct vortices patterns and anomalous criticality. 

\textit{Hamiltonian --}
We study a quantum Rabi zigzag chain under a synthetic magnetic field, realized by arranging two species of coupled cavities in a one-dimensional array, as shown in Fig.~\ref{model}(a). Each upward and downward triangular plaquette is threaded by a staggered magnetic flux, implemented by assigning a photon hopping phase $\theta$ to odd cavities and $\pi-\theta$ to even cavities. Experimental implementation of our system is porposed ( see Supplementary Material (SM)). The Hamiltonian is given by
\begin{eqnarray}\label{hamltonian}
H_{RZ}&=&\sum_{n=1}^{N}H_{R,n}-J_{1}(a_{n}^{\dagger
}a_{n+1}+a_{n+1}^{\dagger
}a_{n})\nonumber\\
&-&J_{2}(-1)^{n}[e^{i\theta }a_{n}^{\dagger
}a_{n+2}+e^{-i\theta }a_{n+2}^{\dagger
}a_{n}],
\end{eqnarray}
where $J_1>0$ represents the photon hopping between nearest-neighbor (NN) cavities of different species, and $J_2>0$ denotes the coupling for next-nearest-neighbors (NNN). Modifying $J_2(-1)^ne^{i\theta}$ in the even or odd chain leads to a sign alternation based on $n$, generating effective magnetic fluxes of $\theta$ and $\pi-\theta$ in  adjacent plaquettes. The artificial magnetic field is realized by tuning the phase lag between the time-varying $J_2(t)$ (see SM~\cite{supple}). Each cavity couples to a two-level atom, characterized by the quantum Rabi Hamiltonian $H_{R,n}=\frac{\Delta }{2}\sigma _{z}+\omega a_n^{\dagger }a_n+g\left( a_n^{\dagger
}+a_n\right) \sigma _{x}$,
where $a_n^{\dagger }$ $\left( a_n\right) $
is the creation (annihilation) operator of the single-mode cavity with frequency $\omega $, and $\sigma _{k}(k=x,y,z)$ are the Pauli matrices. $\Delta $ is the atom transition frequency, and $g$ is the cavity-atom coupling strength. The scaled coupling strength is denoted as  $g_{1}=g/\sqrt{\Delta \omega }$. Such a finite-component system exhibits the superradiant phase transition in the infinite frequency limit $\Delta/\omega\rightarrow\infty$~\cite{Ashhab2013,hwang2015,liu2017,chen2020,chen2021,NCcai2021}.

Applying a unitary transformation $U=\prod_{n=1}^{N}\mathtt{exp}[-ig\sigma
_{y}^n(a_{n}^{\dagger }+a_{n})/\Delta ]$, the effective Hamiltonian projected onto the ground state of the atom $|\downarrow\rangle$ is $H_{\text{RZ}}^{\downarrow } =\sum_{n=1}^{N}\omega a_{n}^{\dagger }a_{n}-\omega g_1^{2}\left( a_{n}^{\dagger }+a_{n}\right)^{2}-[J_{1}a_{n}^{\dagger }a_{n+1} +J_{2}(-1)^{n}e^{i\theta }a_{n}^{\dagger }a_{n+2}+h.c.]$, where high-order terms and the constant are neglected.
Due to the alternating sign present in the NNN hopping strength, we perform the Fourier transform by defining $a_{k}^{\dagger}=\sum_{n=even}e^{-ink}a_{n}^{\dagger}/\sqrt{N}$ and $b_{k}^{\dagger}=\sum_{n=odd}e^{-ink}a_{n}^{\dagger}/\sqrt{N}$ with the momentum $k=2n\pi/N$. The Hamiltonian becomes $H_{\text{RZ}}^{\downarrow }(k)=\sum_{k}\omega_{k+}a_k^{\dagger}a_k+\omega_{k-}b_k^{\dagger}b_k-\omega g_1^2(a_k^{\dagger}a_{-k}^{\dagger}+a_ka_{-k}+b_kb_{-k}+b_k^{\dagger}b_{-k}^{\dagger}) -2J_1\texttt{cos}k(a_k^{\dagger}b_k+b_k^{\dagger}a_k)$, where $\omega_{k,\pm} = \omega(1 - 2g_1^2) \pm 2J_2\cos(\theta - 2k)$. We diagonalize the Hamiltonian as $H_{\text{RZ}}^{\downarrow }(k)=\sum_{k}\epsilon _{+}(k)a_{k}^{\dagger }a_{k}+\epsilon _{-}(k)b_{k}^{\dagger }b_{k}$,  where the energy bands $\epsilon_{\pm}(k)$ dependent on $k$. 
 At $\theta=\pi/2$, $\epsilon_{\pm}(k)$ is analytically given as 
\begin{eqnarray}\label{excitation}
\epsilon_{\pm}^{2}(k)&=&\omega^2(\frac{1}{4}-g_1^2)+J_1^2\mathrm{cos}^{2}k+J_2^2\mathrm{sin}^{2}k\nonumber\\
 &\pm&\omega\sqrt{J_1^2(1-2g_1^2)\mathrm{cos}^{2}k+J_2^2(1-4g_1^2)\mathrm{sin}^{2}k}.
\end{eqnarray}

Take $N=6$ for example. The low excitation energy $\epsilon_{-}({k})$ exhibits two degenerate minima at $k=\pm\pi/3$ with a small $J_1/J_2$ ratio, while it shifts to one minimum at $k=0$ as the ratio increases (see SM).  The vanishing of $\epsilon_{-}({k})$ leads to a critical $k$-dependent coupling strength
\begin{equation}\label{gc}
g_{1c}(k)=\sqrt{\frac{\omega^2-4(J_1^2\mathrm{cos}^{2}k+J_2^2
\mathrm{sin}^2k)}{4\omega(\omega+2J_1\mathrm{cos}k)}}.
\end{equation}
Fig.~\ref{model}(b) displays the ground-state phase diagram at $\theta=\pi/2$. For a weak coupling $g_1$, the system is in the normal phase (NP), characterized by  zero excitation number. As $g_1$ increases, the system undergoes second-order phase transitions from NP to two diﬀerent distinct superradiant phases in Fig.~\ref{model}(b). The phase boundaries are marked by critical lines $g_{1c}(k=0)$ and $g_{1c}(k=\pm\pi/3)$, which join at a critical hopping ratio
\begin{equation}\label{TP}
(J_1/J_2)_c= [\sqrt{\omega^2+12J_2^2}-\omega]/2J_2.
\end{equation}
The critical value $(J_1/J_2)_c$  marks a triple point (TP), where three quantum phases coexist.

\textit{Superradiant phases --} When the coupling strength exceeds $g_{1c}$, the cavity field is macroscopically populated and the system transitions into superradiant phases. In this case, we shift the bosonic operator $a_{n}\rightarrow \tilde{a}_{n}=a_{n}+\alpha _{n}$ with a complex $\alpha_{n}=A_{n}+iB_{n}$. The local displacement $\alpha_n$ can be mapped to classical XY spins $S_n=(S_n^x,S_n^y)$ using the Holstein-Primakoff transformation~\cite{PhysRevLett.129.183602}(see SM~\cite{supple}). 
The lower-energy Hamiltonian projected to the ground state of the atom is obtained (see SM~\cite{supple}) 
\begin{eqnarray}\label{SRham}
H_{\text{RZ}}^{SR}&=&\sum_{n=1}^{N}\omega \tilde{a}_{n}^{\dagger }\tilde{a}_{n}-%
\frac{\lambda _{n}^{2}}{\Delta _{n}}\left( \tilde{a}_{n}^{\dagger }+\tilde{a}_{n}\right)^{2}
-[J_1\tilde{a}_{n}^{\dagger }\tilde{a}_{n+1} \nonumber\\
&+&J_2(-1)^ne^{i\theta}\tilde{a}_{n}^{\dagger }\tilde{a}_{n+2}+h.c.]+E_g,
\end{eqnarray}
where $\lambda_n=g\Delta/\Delta _{n}^{\prime }$ is the effective coupling strength and $\Delta _{n}^{\prime }=\sqrt{\Delta^2+16g^2A_n^2}$ is the renormalized frequency of the atom. 
The ground-state energy is obtained in terms of $\alpha_n$ as $E_g=\sum_{n=1}^{N}\omega (A_n^2+B_n^2)-\frac{1}{2}\sqrt{\Delta^2+16g^2A_n^2}+E_{\text{NN}}+E_{\text{NNN}}$, where the NN and NNN energy terms are $E_{\text{NN}}=\sum_{n=1}^{N}-2J_{1}(A_nA_{n+1}+B_nB_{n+1})$ and $E_{\text{NNN}}=\sum_{n=1}^{N}-2J_{2}(-1)^{n}[\mathrm{cos}\theta(A_nA_{n+2}+B_nB_{n+2})
+\mathrm{sin}\theta(B_nA_{n+2}-B_{n+2}A_n)]$, respectively. Minimizing $E_{\text{NN}}$ results in the shift of the NN cavity  displacement $\alpha_n$ along the $A_n$ or $-A_n$ real axis, resembling ferromagnetic spin alignment. Similarly, the first term in $E_{\text{NNN}}$ induces ferromagnetic or antiferromagnetic displacement dependent on the sign of $-J_2\cos\theta(-1)^n$, while the second term favors a complex displacement $\alpha_n$, resulting in noncollinear displacements. The relative strength of the first and second terms in $E_{\text{NNN}}$ is modulated by $\theta$.

\begin{figure}[tbp]
\includegraphics[scale=0.16]{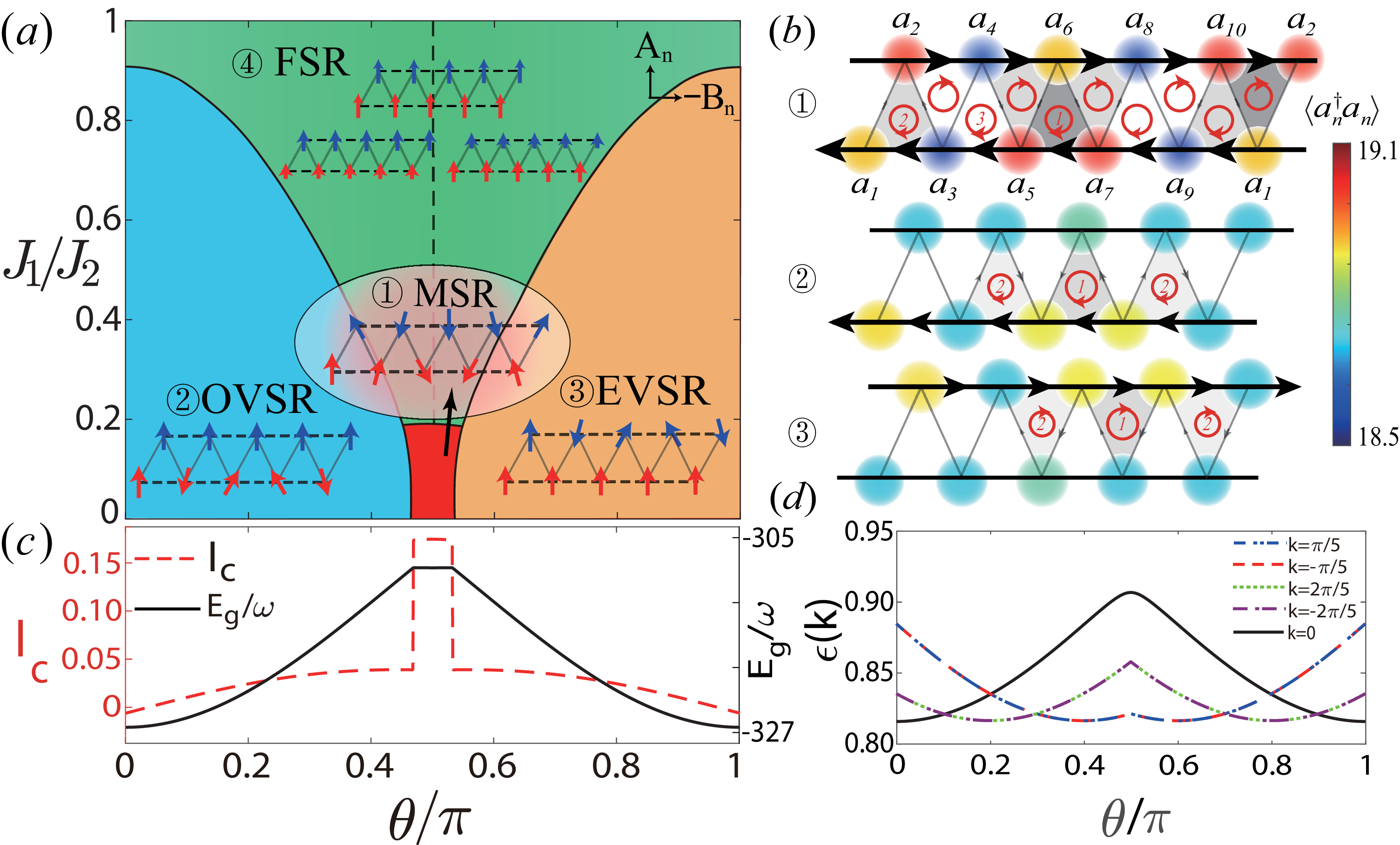}
\caption{(a) Phase diagram in the superradiant regime ($g_1=0.7>g_{1c}$) for $N=10$ cavities.  Phase boundary in solid line separates the MSR, FSR and E(O)VSR phases. Configurations of $\alpha_n$ are marked with red (blue) arrows  in ($A_n$,$-B_n$) plane. (b) Average photon number $\langle a_n^{\dagger}a_n\rangle$  for each phase marked in (a) (1-3).  Three distinct vortex pairs are indicated by gray shading, and the arrow thickness indicates the magnitude of the current. (c) Chiral current $I_C$ (red dashed line) and ground-state energy $E_g/\omega$ (black solid line) versus $\theta$ for $J_1/J_2=0.05$. (d) Excitation energies in Eq.(\ref{excitation}) versus $\theta$ for $g_1=0.1<g_{1c}$ and $J_1/J_2=0.1$. }
\label{phasediagram}
\end{figure}

The displacement $\alpha_n$ can be analytically solved by minimizing the ground-state energy $E_g$ (see SM~\cite{supple}). Fig.~\ref{phasediagram}(a) shows the local displacement configurations along the real $A_n$ and imaginary $B_n$ axis in $N=10$ system. The local photon number is given by $\langle a_n^{\dagger}a_n\rangle=|\alpha_n|^2$. Distinguished superradiant phases are identified  by tuning $\theta$ and $J_1/J_2$. 

(i) Ferromagnetic superradiant (FSR) phase : For strong interspecies hopping above the critical value $(J_1/J_2)_c$, the system exhibits a FSR phase at $k=0$ in Fig.~\ref{model}(b). The dominant NN coupling energy $E_{\text{NN}}$ yields a real $\alpha_n$ of the same sign, exhibiting a ferromagnetic-like configuration. Remarkably, the average photon numbers between odd and even cavities exhibit an asymmetric distribution for $\theta \neq \pi/2$ (see SM), which is distinct from the previous symmetric FSR phase~\cite{PhysRevLett.129.183602}.  It attributes the alternating magnetic flux $\theta$ and $\pi-\theta$ in adjacent plaquettes for $\theta \neq \pi/2$, unlike the symmetric distribution at $\theta=\pi/2$.

Below $(J_1/J_2)_c$, the dominant coupling energy within the same species, $E_{\text{NNN}}$, results in three distinct superradiant phases by tuning $\theta$:

(ii) Meissner superradiant (MSR) phase : For $\theta_{c1}<\theta<\theta_{c2}$ near $\pi/2$, the second term of $E_{\text{NNN}}$, influenced by $\mathrm{sin}\theta$, dominates and induces  the MSR phase. Specifically, at $\theta=\pi/2$, the MSR phase is characterized by the closing of the excitation gap at momentum $k=\pm \pi/5$ as shown in Fig.~\ref{phasediagram} (d).  The corresponding local displacement $\alpha_n$ is complex and displays a noncollinear pattern within the $(A_n,-B_n)$ plane, akin to the chiral spin magnetization in the $x-y$ plane.  An abrupt change in ground state energy $E_g$ at $\theta_{c1}$ and $\theta_{c2}$ in Fig.~\ref{phasediagram}(c) indicate first order phase transitions from the MSR to distinct superradiant phases. 

\begin{figure}[tbp]
\includegraphics[scale=0.145]{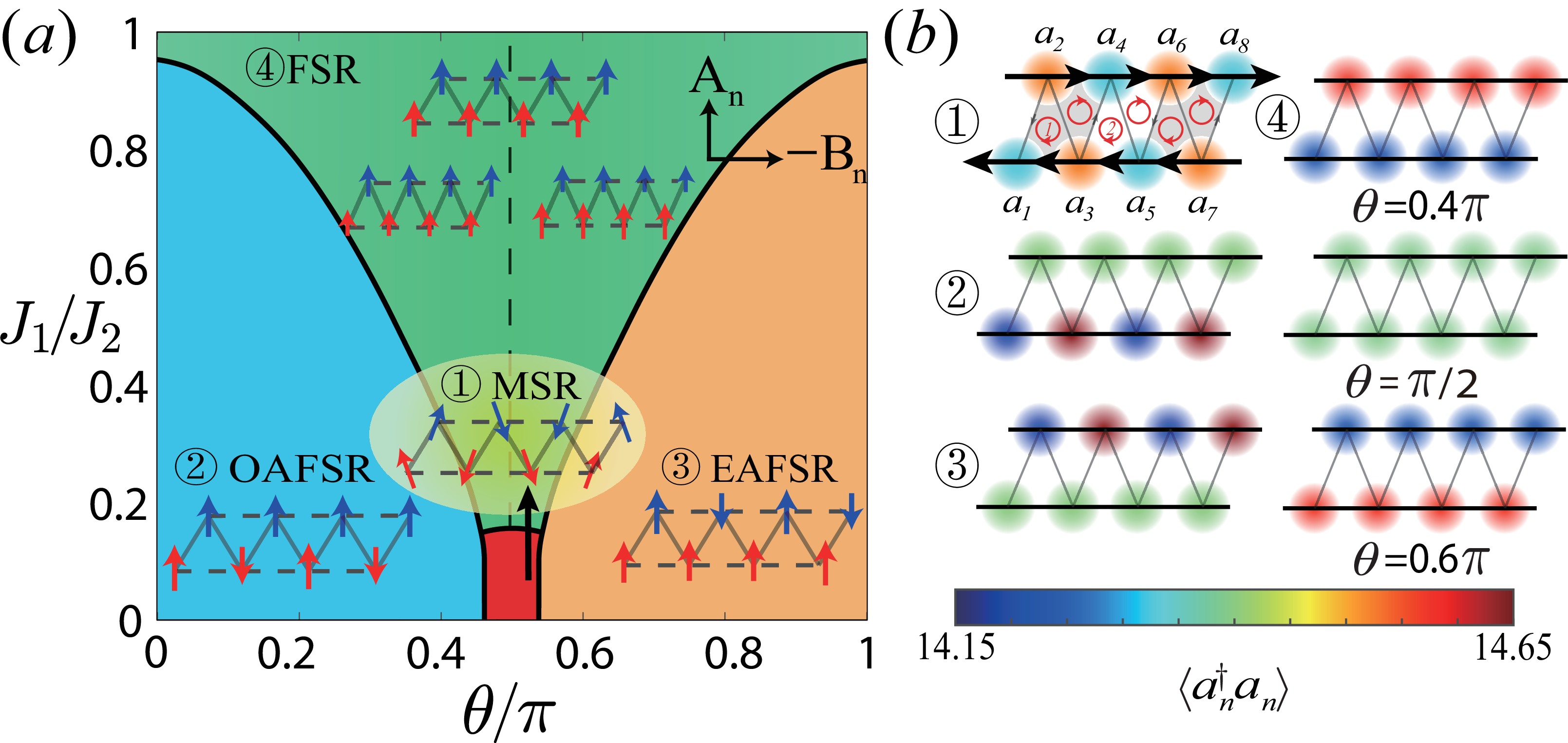}
\caption{(a) Phase diagram in the superradiant regime for $N=8$ cavities. (b)  Currents and average photon number $\langle a_n^{\dagger}a_n\rangle$  for each phase marked in (a) (1-4).  Two distinct pairs of vortex currents are represented by shades of gray. Symbols and parameters align with those in Fig.~\ref{phasediagram}.}
\label{N8PHASE}
\end{figure}

(iii) Odd-vortex superradiant (OVSR) phase : When the flux is small  $\theta<\theta_{c1}<\pi/2$, the system enters a OVSR phase with lower excitation at $k=\pm2\pi/5$ shown in Fig.~\ref{phasediagram}(d).  Odd cavities where $-J_2 \cos \theta (-1)^n>0$ induce frustrated antiferromagnetic patterns due to geometric frustration on both  chain sides with $N/2=5$ cavities.  A cavity consistently exists where $\alpha_n$ remains real, while the remaining $\alpha_n$ values are complex in in Fig.~\ref{phasediagram} (a). This situation is analogous to the frustration observed in antiferromagnetic spins  in a triangular pattern. Conversely, for even cavities, $\alpha_n$ is approximately real and has a uniform sign, indicating a ferromagnetic arrangement.

(iv) Even-vortex superradiant (EVSR) phase : When the magnetic flux is strong with $\theta>\theta_{c_2}>\pi/2$, $\alpha_n$ for the even cavities exhibits a frustrated antiferromagnet pattern, whereas the odd chain aligns ferromagnetically.  Fig.~\ref{phasediagram} (c) illustrates a first-order phase transition from the Meissner phase to two vortex phases characterized by a sudden change in ground state energy,  in analogy to the Meissner effect observed in superconductors.

\begin{figure}[tbp]
\includegraphics[scale=0.37]{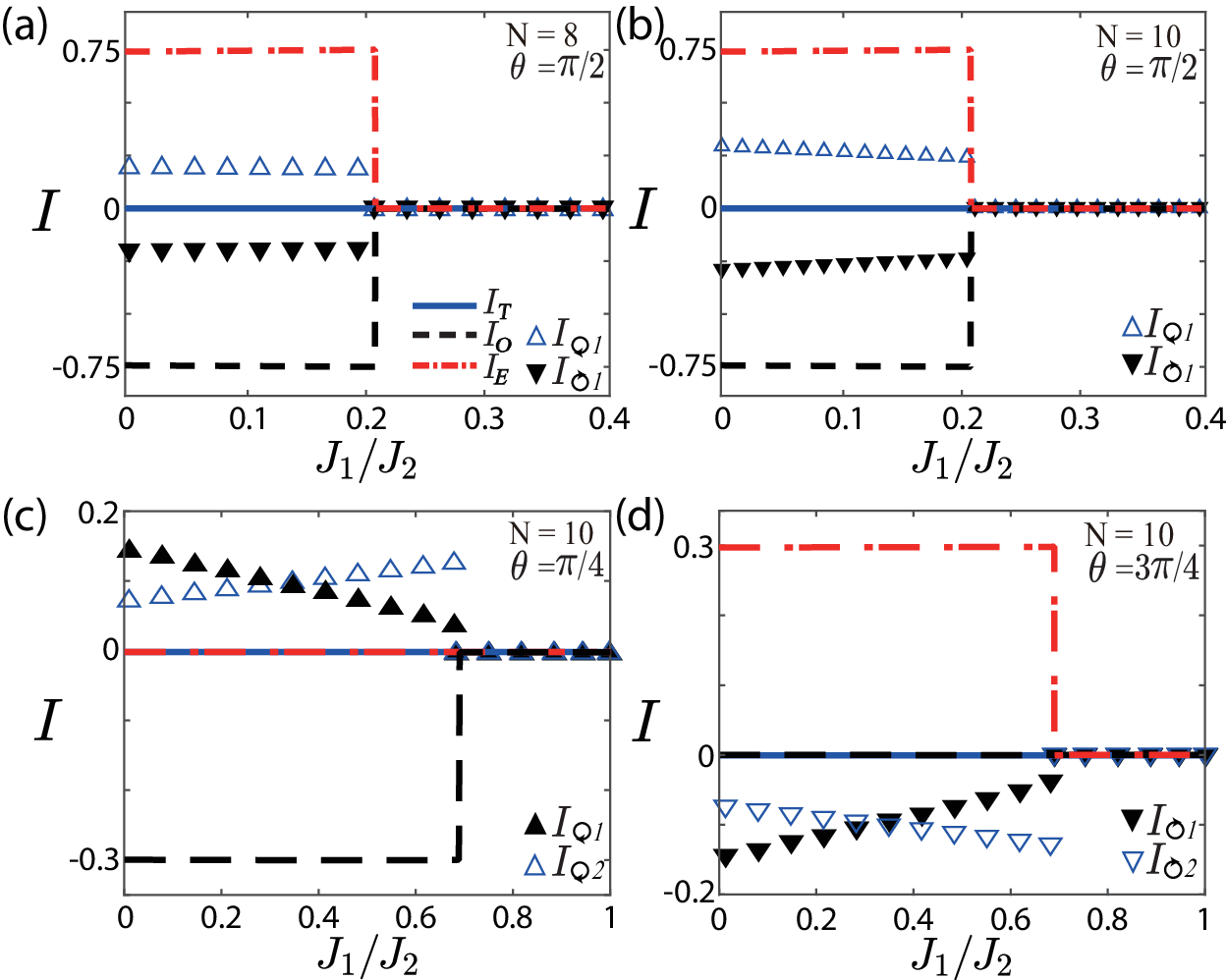}
\caption{Edge currents in odd (even) cavities $I_{O\texttt{(E)}}$ (black dashed (red dotted) lines) and current along the chain $I_\texttt{T}$ (blue solid line ) against $J_1/J_2$ for the MSR-FSR transition with $N=8$ (a) and $N=10$ (b), and for OVSR-FSR (c) and EVSR-FSR transitions (d)  with $g_1=0.7$.  $I_{\circlearrowright1(2)}$ and $I_{\circlearrowleft1(2)}$ in triangle symbols mark distinct vortex currents in triangle plaquettes, with arrows showing circulation direction.}
\label{CURRENT}
\end{figure}

We now examine the superradiant phases in the absence of geometric frustration in $N=8$ system. Each chain side comprises 4 unfrustrated cavities, contrasting with the frustrated geometry of the $N=10$ system. Here, the FSR maintains the real $\alpha_n$ configuration with an asymmetric photon distribution  in Fig.~\ref{N8PHASE}(b), but the MSR phase demonstrates unique chiral local displacement patterns with stable, site-specific spin orientations, as depicted in Fig.~\ref{N8PHASE} (a). Unlike the OVSR phase, the $N=8$ system favors a stable antiferromagnetic configuration in odd cavities, forming an odd-antiferromagnetic superradiant (OAFSR) phase. Similarly, an even-antiferromagnetic superradiant (EAFSR) phase supplants the EVSR phase.

\textit{Chiral edge current --}We investigate photon currents within zigzag cavity loops to examine chirality in anomalous superradiant phases. Analogous to the classical continuity equation, the photon current operator along the chain is $\hat{I_\texttt{T}}=\sum_{n=1}^{N} iJ_1( a_{n}^{\dagger}a_{n+1}-a_{n+1}^{\dagger}a_{n})$, which is approximated as $I_\texttt{T}=\sum_{n=1}^{N} -2J_1\text{Im}(\alpha_{n}^{*}\alpha_{n+1})$ in the mean-field approximation. Similarly, $I_{\nu}$ with $\nu=(E,O)$, representing currents on even or odd  side are described by $I_{\nu}=\sum_{n=\nu}-2J_2(A_nB_{n+2}-B_nA_{n+2})$.  The chiral current is given by $I_\texttt{C}=I_O-I_\texttt{E}$.  Fig.~\ref{phasediagram}(c) displays the chiral current across the Meissner-to-vortex phase transition. As $\theta$ increases, $I_\texttt{C}$ first grows  in the OVSR phase, sharply jumps to a maximum at the critical flux $\theta_{c1}$ in the MSR phase, then decreases in the EVSR phase $\theta>\theta_{c2}$. 
 
Fig.~\ref{CURRENT} illustrates chiral edge currents $I_{\texttt{E(O)}}$ along each side of the chain by varying $J_1/J_2$. At $\theta=0$ or $\pi$, the current is absent. When $\theta=\pi/2$ in the MSR phase, the currents in the even and odd chains flow in opposite directions, meaning $I_\texttt{E} \times I_O < 0$. The $N=8$ system hosts two pairs of vortices in adjacent triangle plaquettes in Fig.~\ref{N8PHASE}(b). Each exhibits counter-circulating currents  ($I_{\circlearrowright1}$ and $I_{\circlearrowleft1}$)  that cancel to yield a vanishing net current  $I_{\texttt{T}} = 0$ in Fig.~\ref{CURRENT}(a) . Notably,  $N=10$ system exhibits a non-uniform vortex structure featuring three distinct  pairs of vortices in Fig.~\ref{phasediagram}(b), arising from frustrated photon distribution. However, the overall current remains zero $I_{\texttt{T}} = 0$  in a sequence of cancellation of antiparallel vortex pairs in Fig.~\ref{CURRENT}(b).  The chiral edge current resembles surface currents observed in the Meissner effect of superconductors. Notably, these edge currents remain robust even under controlled inhomogeneities, as long as the cavities preserve photon-configuration symmetry, maintaining antiparallel vortex pairs (see SM~\cite{supple}). Consequently, both the robust antiparallel vortex pairs and the persistent edge currents exhibit a quantum-Hall-like robustness, despite being non-topological ~\cite{PhysRevA.89.023619,PhysRevLett.105.255302}. In the FSR phase, the currents abruptly disappear above the critical threshold $(J_1/J_2)_c$. In the OVSR phase ($\theta=\pi/4$), two vortex currents $I_{\circlearrowright1(2)}$ localize exclusively in upward triangular plaquettes in Fig.~\ref{CURRENT}(c), while a chiral current emerges exclusively in the odd chain $I_O<0$. Conversely,  the EVSR phase exhibits a chiral current solely in the even chain $I_\texttt{E}>0$, with two vortex currents  localized in downward triangular plaquettes in Fig.~\ref{CURRENT}(d).

\begin{figure}[tbp]
\includegraphics[scale=0.23]{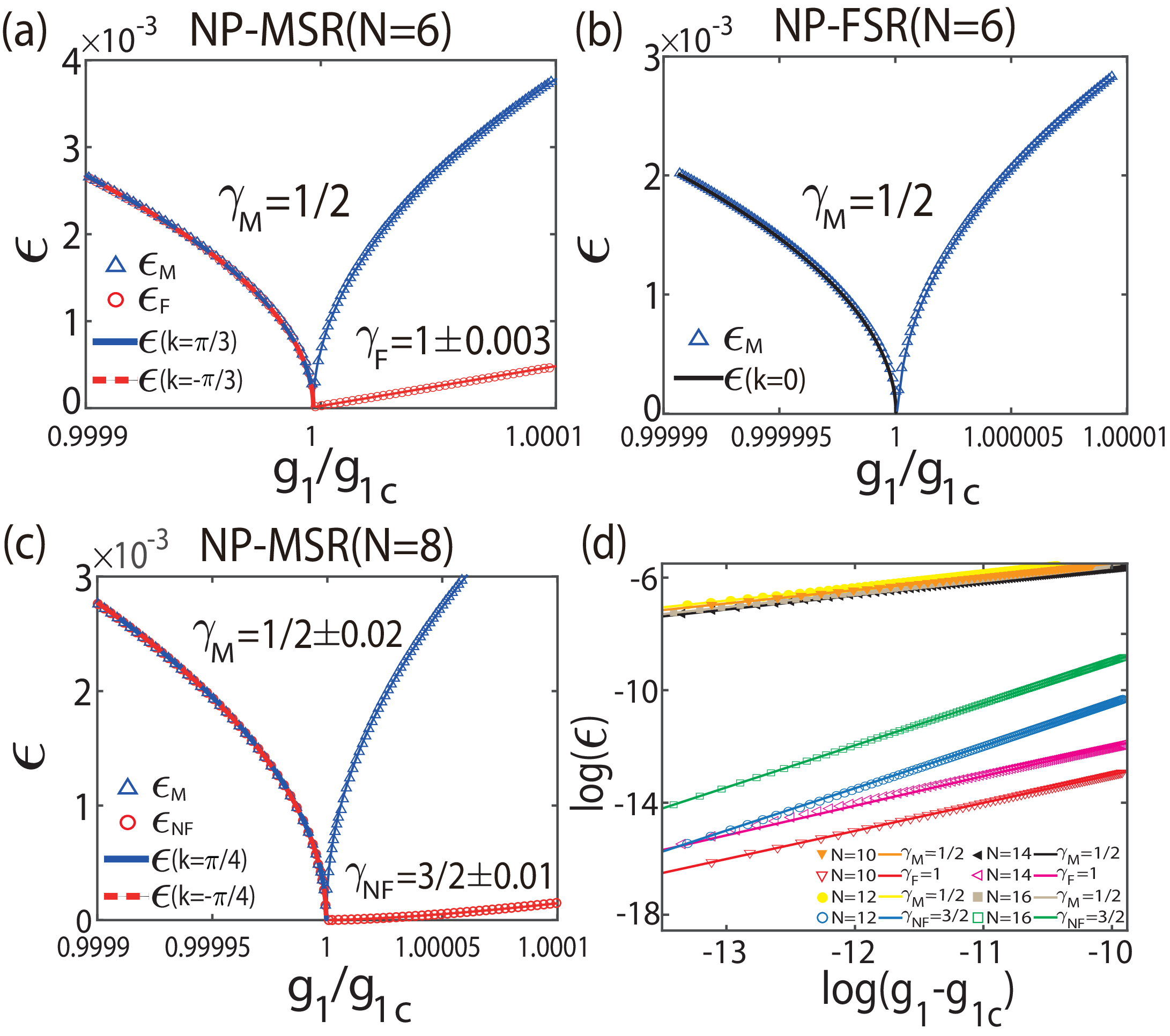}
\caption{Critical excitation energies follow $\epsilon$ versus $g_1/g_{1c}$ for the NP-MSR (a) and NP-FSR (b) transitions ($N=6$), and for the NP-MSR transition ($N=8$) (c) with distinct power laws with exponents $\gamma_\texttt{{M}}=1/2$ ,  $\gamma_\texttt{{F}}=1$ and  $\gamma_\texttt{{NF}}=3/2$. Shapes represent numerical results, which fit well with  to analytical $\epsilon(k)$ ( lines) from Eq.(\ref{excitation}) for $g_1<g_{\texttt{1c}}$. (d) Excitation energy scaling exponents for $N=10,12,14,16$ in the NP-MSR transition, with lines showing a power law. }\label{scaling}
\end{figure}

\textit{Excitation energies --} We analyze excitation energy behavior close to phase transitions. Using bosonic operators $\beta=\{b_{n}^{\dagger }, b_{n}\}$ as a linear combination of $\alpha
=\{\tilde{a}_{n}^{\dagger },\tilde{a}_{n}\}$, the lower-energy Hamiltonian in the superradiant phases can be diagonalized as $H_{\text{RZ}}^{SR}=2\sum_{n=1}^N\epsilon_nb_n^{\dagger}b_n+(\epsilon_n-\omega)/2$ with the excitation spectrum $\epsilon_n$ (see SM~\cite{supple}). 
Near the critical point $g_{1c}$, the lower excitation energy $\epsilon$ vanishes as $\epsilon\propto|g_1-g_{1c}|^{\gamma}$ with a critical exponent $\gamma$.  For the NP-FSR transition, the lower excitation branch $\epsilon_{\texttt{M}}$ of a mean-field mode exhibits critical scaling with exponent $\gamma_{\texttt{M}}=1/2$ in Fig.~\ref{scaling} (b), which is the same as the coventional superradiant transition in the Dicke model~\cite{PhysRevLett.92.073602,PhysRevA78}.  However, the NP-MSR transition for $N=6$ shows two distinct modes: a mean-field mode $\epsilon_{\texttt{M}}$ with $\gamma_{\texttt{M}}=1/2$, and a frustrated mode $\epsilon_{\texttt{F}}$ with nonsymmetric exponents $\gamma_{\texttt{M}}=1/2$ and $\gamma_{\texttt{F}}=1$ below and above the transition in Fig.~\ref{scaling}(a). Conversely, the $N=8$ system  exhibits a non-frustrated mode $\epsilon_{\texttt{NF}}$  with asymmetric exponents: $\gamma_{\texttt{M}}=1/2$ and $\gamma_{\texttt{NF}}=3/2$.    As $N$ increases,  Fig.~\ref{scaling}(d) shows three distinct critical scaling behaviors emerging simultaneously:  {\color{blue}the mean-field exponent $\gamma_{\texttt{M}}=1/2$, the frustrated-mode exponent $\gamma_{\texttt{F}}=1$ for odd-$N/2$ chains ($N=6,10,14$), and the non-frunstrated one $\gamma_{\texttt{NF}}=3/2$ for even-$N/2$ chains ($N=8,12,16$).}  The unususal exponents  $\gamma_{\texttt{F(NF)}}$ stem from  geometry frustrations and the magnetic flux, revealing distinct universality classes of the MSR phase transitions.

\textit{Conclusion --} 
We have explored the Meissner-like photon chiral edge currents in the two-species Rabi cavities coupled system subjected to a staggered magnetic field.  Our findings demonstrate how Meissner and vortex phases emerge alongside photon superradiance under tunable artificial gauge fields and photon hoppings. Geometric frustration leads to unique vortex patterns and unconventional criticality, uncovering diverse universality classes in phase transitions. This optical platform allows highly controlled quantum simulation of persistent chiral currents and photon vortices, potentially connecting to quantum Hall-like edge states and its related applications.

\textit{Acknowledgments--}
We thank Han Pu for insightful discussions. This work was supported by National Natural Science Foundation of China Grant No.12475013, No. 12474140, and No. 12347101.

\bibliography{refs}{}

\end{document}